# Pitfalls in Determining the Electrical Bandwidth of Non-Ideal Nanomaterials for Photodetection

Christine Schedel[1,a], Fabian Strauß[1,2,a], Marcus Scheele[1,2,*]

[1] Institute for Physical and Theoretical Chemistry, University of Tübingen, 72076 Tübingen, Germany.

[2] Center for Light-Matter Interaction, Sensors and Analytics LISA+, University of Tübingen, 72076 Tübingen, Germany.

* Corresponding author: marcus.scheele@uni-tuebingen.de

[a] These authors contributed equally.

## Abstract
The electrical 3 dB bandwidth is regularly used as a measure for the response speed of a photodetector and is estimated via various approaches in literature, ranging from direct measurements to gauged values via approximations. Great care must be taken when comparing these 3 dB bandwidths, since approximations are only strictly valid for ideal circuits. This paper demonstrates that, for typical photodetectors based on new emerging nanostructured materials, namely quantum dots and transition metal dichalcogenides, the bandwidth can deviate up to $10^3$ depending on the chosen approach for the bandwidth specification.

## Introduction
For the development of new materials and devices for time-resolved (opto-)electronics, the analysis in the time- or frequency-domain is a pivotal task. Nearly a hundred years ago, a first connection between those two domains was established by Küpfmüller's uncertainty principle,[1] which laid the foundation for connecting the electrical bandwidth to the time-response in telecommunication. In data transmission and receiving, the bandwidth is a crucial measure for the maximum frequency of device operation under which two data points can still be distinguished. Traditionally, the 3 dB bandwidth is taken, which is the frequency at which the power of the response has decreased to fifty percent[2,3]. The main approaches for determining the electrical 3 dB bandwidth of a photodetector are either direct measurements in the frequency domain[4,5] or measuring the time-resolved photocurrent *vs.* varying laser repetition rates[6–9]. If only one repetition rate is chosen, the impulse photoresponse ($f(t)$) is regularly used to calculate the power spectrum ($P(\omega)$) via fast Fourier transformation (FFT)[10–19]:

$$P(\omega) = |\text{FFT}(f(t))|^2 \qquad (1)$$

Upon normalization against the power under steady state conditions ($P_1$), the electrical bandwidth is obtained according to  dB $= 10\log_{10}(P(\omega)/P_1)$.

Another, more convenient approach involves the analysis of the response time ($\tau$) or the rise-time ($t_{rise}$) obtained by time-resolved square pulse-[7,8,20] or sinusoidal-[20–22] illumination of the device. However, these techniques require a conversion formula to approximate the 3 dB bandwidth $f_{3dB}$, and by far the most commonly used one is [5–7,20,23–30]:

$$f_{3dB} = 0.159/\tau = 0.35/t_{rise} \qquad (2)$$

We stress that this approximation only holds true for ideal circuits consisting of an RC-device behaving like a first order pole low-pass filter[31,32]. Other approximations used range from $f_{3dB} = 1/t_{rise}$[33–35] up to $f_{3dB} = 2.2/t_{rise}$[36]. In addition, Equation (2) is also applied to infer the 3 dB bandwidth from the impulse response of a material by inserting the fall time in place of the rise time into Equation (2)[37].

In light of the growing attention devoted to nanomaterials and their photodetecting properties,[38,39] we note that the above-mentioned approximations are frequently applied, although many nanostructured photodetectors are far from representing ideal circuits and the required first order pole low-pass filter characteristics. This is likely to introduce inconsistencies in the reported 3 dB bandwidths with potentially misleading conclusions for the comparison of studies that apply different methodologies for determining $f_{3dB}$. The purpose of this work is to evaluate the magnitude of the deviations in $f_{3dB}$ for typical nanomaterials when the ideal circuit approximation is applied compared to the exact bandwidth determinations outlined above.

Specifically, we investigate a commercial silicon p-i-n photodiode for reference, two detectors based on quantum dots (QDs) and two devices made from the transition metal dichalcogenides (TMDCs) $MoS_2$ and $WSe_2$. The QD detectors are iodide-capped CdSe QDs crosslinked with the organic dye zinc β-tetraaminophthalocyanine, (hereafter referred to as CdSe/I⁻/Zn4APc) as well as PbS QDs crosslinked with ethane-1,2-dithiol (PbS/EDT). We use square-pulse measurements for the steady-state and impulse excitation for the non-steady state response. From the latter, we determine $f_{3dB}$ according to Equation (1). We compare this value to the approximated bandwidths according to Formula (2).

## Experimental Section

**Detectors.** The commercially available photodiode *BPW 34* (expected rise time: 20 ns for 5 V reverse bias, 50 Ω load resistance) by *Osram Opto Semiconductors*[40] was investigated.

For the PbS/EDT photodetector the particles were synthesized after a route of Weidman et al.[43] and the ligand exchange was performed according to Luther et al.[44]. The films were prepared on a glass substrate with interdigitated electrodes (4 nm Ti / 20 nm Au, 2.5 μm × 1 cm, finger width 20 μm, film height) as reported elsewhere[45]. In short: 7.5 ml oleylamine and 0.04 g sulfur were heated to 120 °C for 20 min in an oil bath under nitrogen atmosphere before allowed to cool back to room temperature while maintaining the nitrogen flow. For the lead solution 7.5 ml oleylamine and 2.5 g $PbCl_2$ were degassed for 10 min at a Schlenk line (150 mTorr) before heating under oxygen exclusion to 120 °C. The flask temperature was held at 120 °C while 2.25 ml of the previously prepared sulfur solution was injected. Then the reaction was quenched after 6 min by putting the flask into a water bath and adding 20 ml cold hexane, yielding 6 nm sized PbS quantum dots. Device preparation was done in a nitrogen filled glovebox by covering a glass substrate with 75 μl of the PbS hexane solution and spin coating after 30 s for 30 s with a speed of 20 rps. Then, 150 µl EDT acetonitrile solution (5 mM) was added and again spin coated with the same parameters after 30 s. 150 µl acetonitrile were added and after 30 s spin coated (30 s, 20 rps). All three steps were repeated twice, before the device was put under vacuum for half an hour. The film thickness amounted to approximately 50 nm. The detector was examined under vacuum.

For the CdSe/I$^-$/Zn4APc devices, 4.5 ± 0.4 nm sized CdSe QDs were synthesized according to Sayevich et al.[46] In short, 176.0 mg CdO, 8.0 g trioctylphosphine oxide, 8.0 g hexadecylamine, 2.2 mL oleic acid, and 45.8 mL 1-octadecene were kept under vacuum (~10$^{-3}$ mbar) for 2 h and were heated to 300 °C under nitrogen afterwards. A clear solution was formed, then the temperature was reduced and held at 275 °C for 30 min. A solution of 126.0 mg Se in 4 mL trioctylphosphine, 4 mL trioctylphosphine, and 4 mL 1-octadecene was injected into the reaction mixture, then the temperature was increased to 280 °C and kept there for 45 min. Afterwards, the reaction was quenched by a sudden cooling with cold water. The QDs were purified by precipitation with ethanol (twice) and acetone (twice) and redispersed in hexane. For the ligand exchange with iodide, 0.84 mL of a 60.5 mg/mL CdSe solution in hexane was stirred over night with 300 µL of a 1 M solution of NH$_4$I in N-methylformamide and 2.7 mL acetone. The mixture was centrifuged, washed with acetone, and redispersed in 400 µL N-methylformamide. Subsequently, the CdSe/I$^-$/Zn4APc films were prepared as reported elsewhere[47]. In short, 45 µL of an 88 µM QD solution in N-methylformamide and 65 µL of a saturated Zn4APc solution in DMSO were deposited on a glass substrate with interdigitated electrodes on top (4 nm Ti / 20 nm Au, 350 nm × 10 mm, finger width 80 nm). After film formation over-night, excess solvent was spun-off the substrate, the film was washed with acetonitrile to get rid of unbound Zn4APc, and the film was annealed at 190 °C for 30 min. The film was inhomogeneous with up to 3.5 µm thickness, as reported elsewhere[48]. The detector was examined under vacuum.

The WSe$_2$ transition metal dichalcogenide detectors were prepared as reported elsewhere[49] using scotch tape exfoliation of TMDC flakes onto lithographically fabricated electrodes (WSe$_2$: 2 nm Ti / 10 nm Au, MoS$_2$: 4 nm Ti / 20 nm Au, 5 µm × 80 µm, finger width 10 µm, flake thickness WSe$_2$: 30 nm, flake thickness MoS$_2$: 110 nm). In contrast, the MoS$_2$ detectors were fabricated by exfoliating the TMDC first before performing lithography onto the flakes. Prior to evaporation of the electrode material, the flake was etched with a mixture of O$_2$/SF$_6$ plasma (100 W, 25 % O$_2$ / 75 % SF$_6$, 50 mTorr, 15 s) to get an edge-on contact[50,51]. Both TMDC detectors were prepared onto glass substrates and examined under atmosphere.

**Transient photoresponse**. The time-resolved photocurrent measurements were carried out at room temperature in a Lake Shore Cryotronics probe station CRX-6.5K. Steady state photoelectric response was measured using square pulse illumination of the photodetectors. A fast switchable laser driver (FSL500, PicoQuant) was used, operated with a Hewlett Packard 33120A arbitrary waveform generator at 100 Hz. The 635 nm laser diode has a laser rise time of < 0.5 ns and an optical output power of ≤ 12 mW. The impulse photoresponse of the photodetectors was examined with a picosecond pulsed laser driver (Taiko PDL M1, PicoQuant), using a 636 nm laser head with a pulse width of < 500 ps. For 100 kHz repetition rate, 22 µW average optical power was chosen for the nanostructured detectors. The given laser powers were reduced by inefficient coupling into the optical fiber, scattering, decollimation of the beam and due to the different detector sizes, an unfocused beam was used with the laser spot usually being larger than the detector area under study.

The photodetector electrodes were contacted with 50 Ω matched tungsten probes and 40 GHz coaxial cables chosen as short as possible. For the homebuilt detectors, the current was preamplified with a FEMTO DHPCA-100 current amplifier. The photocurrent of all devices was measured with a Zurich Instruments UHFLI lock-in amplifier with a Periodic Waveform Analyzer Function, which averaged the signal from 2 G samples. All signals were background corrected. The maximum time resolution amounts to 600 MHz, which is the signal input limitation of the lock-in amplifier.

For the QD devices the Fourier transform was done after applying zero-padding to mimic a 12.5 kHz measurement for the CdSe and a 250 Hz for the PbS device in order to determine the 3 dB bandwidth.

# Results & Discussion

## Commercial photodiode

The normalized photoresponses of the commercially available silicon p-i-n photodiode BPW 34 by *Osram Opto Semiconductors* are shown in Figure 1. The photoelectric response towards a 635 nm 3 MHz square pulse illumination with 100 ns pulse width under 5 V reverse bias and with a 50 Ω load resistance is given in Figure 1a. Steady state photocurrent is reached and a rise time (10 – 90 %) of 23.8 ns is determined. This matches the rise time of approx. 20 ns given in the data sheet of the diode[40]. We approximate $f_{3dB}$ with 14.7 MHz according to Equation (2).

The response of the detector towards 636 nm 3 MHz impulse illumination is depicted in the inset of Figure 1b, and the bandwidth spectrum is calculated according to Equation (1), shown in Figure 1b. A 3 dB bandwidth of 18.0 MHz can be specified. With the fall time of the impulse response used as an estimate for the rise time of the square pulse response, a bandwidth of 20.7 MHz is expected. The bandwidths determined by the three approaches match closely, indicating a near-ideal circuit, and the diode can be specified with a bandwidth of roughly 17.8 ± 3 MHz. All bandwidth values are depicted in Figure 3.

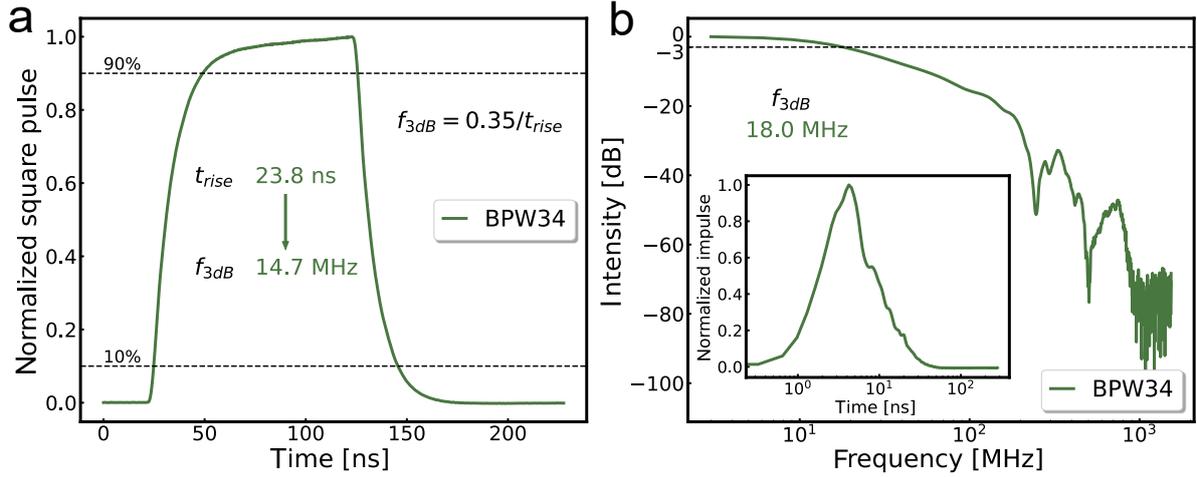

*Figure 1. a) Normalized photoresponse of a BPW 34 photodiode towards a 635 nm 3 MHz square pulse laser. b) Fourier transformed photoresponse of the same diode towards a 636 nm 3 MHz impulse laser. Inset: normalized impulse photoresponse.*

### Photodiodes based on nanostructured materials

The channel lengths for the nanostructured photodetectors are 5 µm for PbS/EDT, 350 nm for CdSe/I$^-$/Zn4APc, 15 µm for MoS$_2$ and 5 µm for WSe$_2$. While we used bottom electrodes for the QD detectors, we exploited both bottom face-on and edge-on electrode geometries for the TMDC based detectors, to enable a better transfer of the charge carriers to the electrode[41]. We worked under low electric fields of 0.40 kV/cm and 0.67 kV/cm for the WSe$_2$, MoS$_2$ detectors, respectively, to avoid breakdown of the devices. For the more stable PbS/EDT and CdSe/I$^-$/Zn4APc devices, we applied 5.00 kV/cm, 85.7 kV/cm, respectively. We measured the QD devices under vacuum due to their susceptibility to oxygen and the more inert TMDC devices under atmosphere. A detailed description of the photodetector preparation is given in the section Materials and Methods. Thus, the choice of nanostructured devices investigated here reflects a large variety of different materials, electrode geometries, preparation procedures, and measurement conditions.

The photoresponses of the four different nanostructured detectors are given in Figure 2, with the measurements performed on QD systems shown in Figures 2a (square pulse), 2b (impulse), and on TMDCs shown in Figures 2c (square pulse), 2d (impulse).

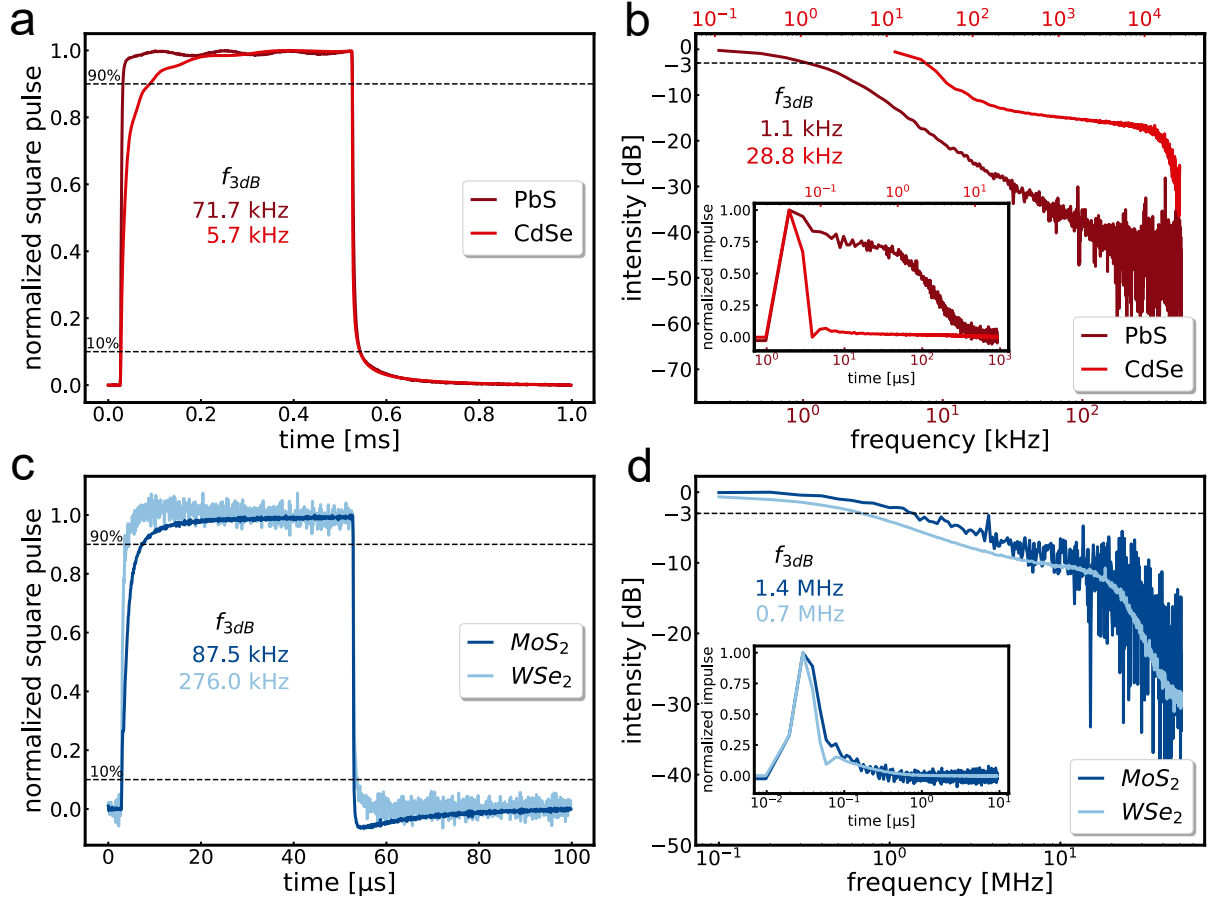

*Figure 2. a) Normalized photoresponses of the four photodetectors towards both, square pulse illumination (a, c) and impulse illumination (b, d), giving the bandwidth spectrum. The two QD systems are shown in a) towards a 635 nm 1 kHz square pulse laser and in b) with the bandwidth spectrum determined via Fourier transformation of the photoresponse of the QDs towards a 636 nm impulse laser with 50 kHz (CdSe) and 1 kHz (PbS) laser repetition rate. The responses of the TMDC systems are shown in c) towards a 635 nm 10 kHz square pulse illumination and in d) together with the bandwidth spectrum for a 636 nm 100 kHz impulse illumination.*

For the PbS/EDT device, we find a rise time of 4.9 µs which translates into a 71.7 kHz 3 dB bandwidth using Equation (2). Compared with the result of 1.1 kHz determined via Fourier transformation of the data in Figure 2b, this seventy-fold difference highlights the poor applicability of the first order pole low-pass filter approximation in this case. Similarly, the CdSe/I$^-$/Zn4APc exhibits a rise time of 61.5 µs, which corresponds to an approximated 5.7 kHz bandwidth. In contrast, the bandwidth determined via Fourier transformation is five times larger with 28.8 kHz. We note that using the impulse fall time for the approximation via Equation (2) agrees perfectly with the result after Fourier transformation for PbS/EDT (1.1 kHz), but with 18.0 MHz it deviates by almost three orders of magnitude for CdSe/I$^-$/Zn4APc.

For MoS$_2$, the approximated bandwidth is 87.5 kHz (Figure 2c), which compares to 1.4 MHz upon Fourier transform of the impulse photoresponse (Figure 2d). Likewise, the approximated bandwidth of the WSe$_2$ device is 276.0 kHz, compared to 0.7 MHz after Fourier transform of the impulse response. When utilizing the impulse response fall time in Equation (2), the bandwidths are estimated with 3.6 MHz for both devices. All bandwidths estimated with Equation (2) or the Fourier transformation of the impulse, Equation (1), are shown in Figure 3 for easy comparability.

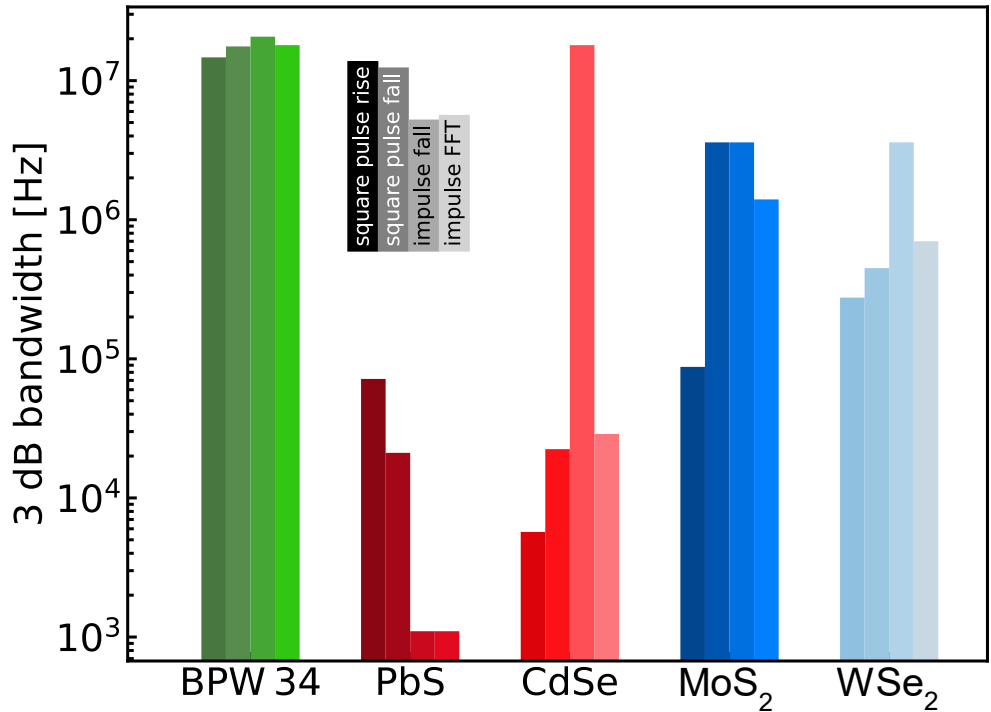

*Figure 3. Overview of all bandwidths derived with either Equation (2): $f_{3dB} = 0.35/t$ using the respective square pulse rise / fall time or the impulse fall time or the bandwidth estimated using the Fourier transformation of the impulse response with Equation (1).*

While the analysis of the commercial photodiode (representing a near-ideal circuit) verifies that all four approaches for determining the 3 dB bandwidth utilized here can lead to good agreement, significant differences are observed for the nanostructured devices. The largest deviations arise for the CdSe detector with a factor of $3 \cdot 10^3$. The PbS detector and the MoS$_2$ detector show deviations of around 65 and 40, respectively, and for the WSe$_2$ device, still a factor of 10 is detectable. We argue that the bandwidth determined via Fourier transformation of the impulse response gives the most accurate result, since it uses a complete signal as basis for the calculation without any approximation. It is evident that the approximation in Equation (2) can lead to drastic over- and underestimations of the 3 dB bandwidth without a clear trend.

We note that the 3 dB bandwidth can depend on the illumination intensity,[36,42] which needs to be considered as an alternative explanation for the different results obtained from the square pulse *vs.* the impulse data. However, two observations speak against a dominant contribution of this dependence: I) We find no significant deviations for the commercial p-i-n photodiode despite using two different laser sources. II) Even the results obtained from the impulse photoresponse with the same illumination intensity yield largely different values as demonstrated for the CdSe/I⁻/Zn4APc detector, where the deviation amounts to a factor of 625.

These large deviations in the 3 dB bandwidth of one and the same photodetector highlight that care must be taken when comparing 3 dB bandwidths of different devices specified via different approaches. We suggest to primarily compare the bandwidth values detected with the identical approach.

# Conclusion

We apply four related methods to determine the electrical 3 dB bandwidth of photodetectors based on CdSe or PbS nanocrystals, $MoS_2$ or $WSe_2$ nanoflakes as well as a commercially available photodiode for reference. While we find similar results with all four methods for the commercial, ideal photodiode, the derived 3 dB bandwidths for the nanostructured, non-ideal photodetectors differ by up to $10^3$ depending on the applied method/approximation. These discrepancies need to be taken into account when comparing non-ideal photodetectors – such as most devices based on nanoparticles - with existing data. We recommend comparing only bandwidths obtained with the same method.

# Acknowledgements

We thank Patrick Michel for the synthesis of the CdSe QDs. Furthermore, we acknowledge Alexander André for the synthesis of the PbS QDs and Pia Kohlschreiber for the PbS/EDT thin film preparation. Financial support of this work has been provided by the European Research Council (ERC) under the European Union's Horizon 2020 research and innovation program (Grant Agreement 802822) as well as the DFG under grant SCHE1905/9-1.

TOC Graphic

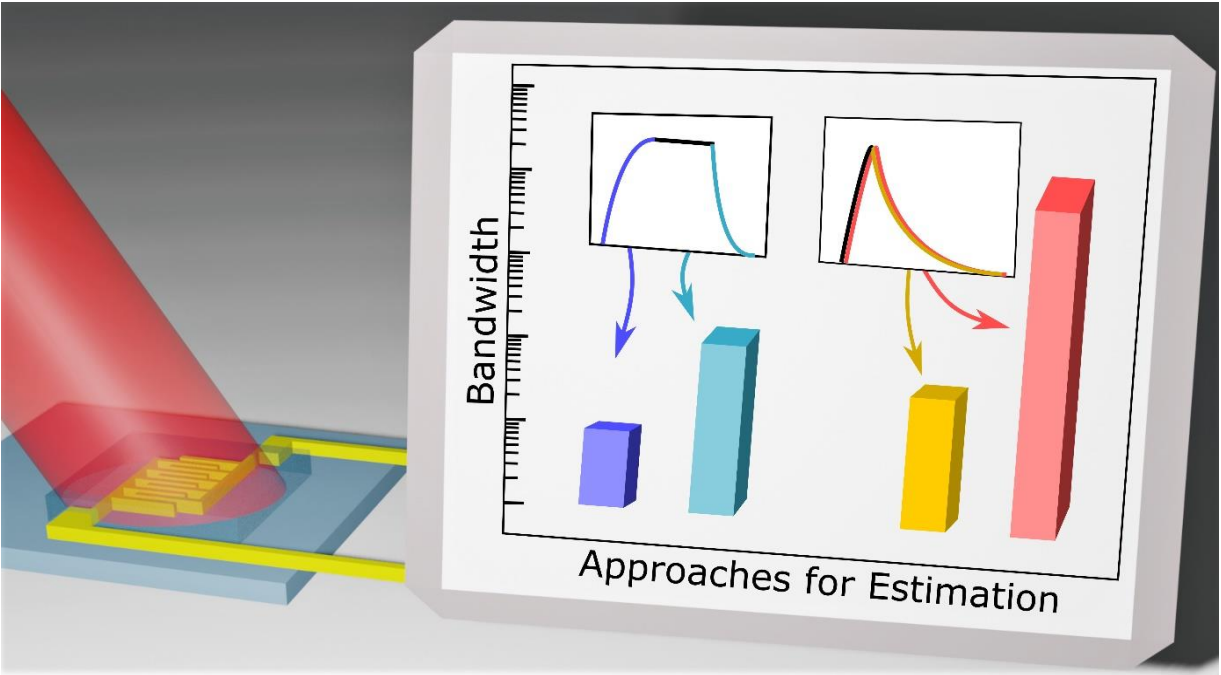